\documentclass[preprint,12pt]{elsarticle}

\usepackage{amssymb}

\journal{Physica C}

\begin{document}

\begin{frontmatter}



\title{Comparatively high in-field critical current in type-II superconductors from heterogeneous columnar pins: a molecular dynamics study}


\author{J. P. Rodriguez}
\author{E. J. Oswald}

\address{Department of Physics and Astronomy, 
California State University at Los Angeles,
Los Angeles, CA 90032}

\begin{abstract}
Theoretical work predicts that the strong dependence of $T_c$ on pure shear strain
within the $a$-$b$ plane of optimally doped YBa$_2$Cu$_3$O$_{7-\delta}$  
results in heterogenous columnar pins of vortex lines 
about dislocation lines
and about nano-columns inclusions 
aligned in parallel to the $c$ axis.  The critical
current of a rigid vortex lattice driven by the Lorentz force in the presence
of such clusters of pin/antipin lines is computed using
two-dimensional (2D) collective pinning theory
and by numerical simulation of the corresponding 2D vortex dynamics.  
Both theory and computer calculation
find that the antipin component of the heterogenous
columnar pins contributes substantially to the net in-field critical current.

\end{abstract}


\end{frontmatter}


\section{Introduction}
\label{intro}
Optimally doped films of the high-$T_c$ superconductor
YBa$_2$Cu$_3$O$_{7-\delta}$ (YBCO) 
grown by pulsed laser deposition (PLD) 
show high critical currents 
that reach a substantial fraction of the theoretical maximum depairing current
in applied magnetic field oriented in parallel to the $c$ axis\cite{review}.
Evidence exists that this is primarily due to 
edge dislocations that thread the film along 
the $c$ axis in PLD-YBCO\cite{klaassen}.
The critical current of YBCO films in $c$-axis aligned applied magnetic field
can also be enhanced substantially by the introduction of BaZrO$_3$ (BZO) 
or of BaSnO$_3$ (BSO) material that naturally form nano-rods 
along the $c$ axis during the film growth process\cite{bzo}\cite{goyal}\cite{bso_07}.
Such increases in the in-field performance of the critical current in YBCO
films by columnar microstructures are very important to the
development of second-generation high-$T_c$ superconducting wire technology.
Achieving a fundamental understanding of how they occur is therefore desirable.

\begin{figure}
\includegraphics[scale=1.30, angle=0]{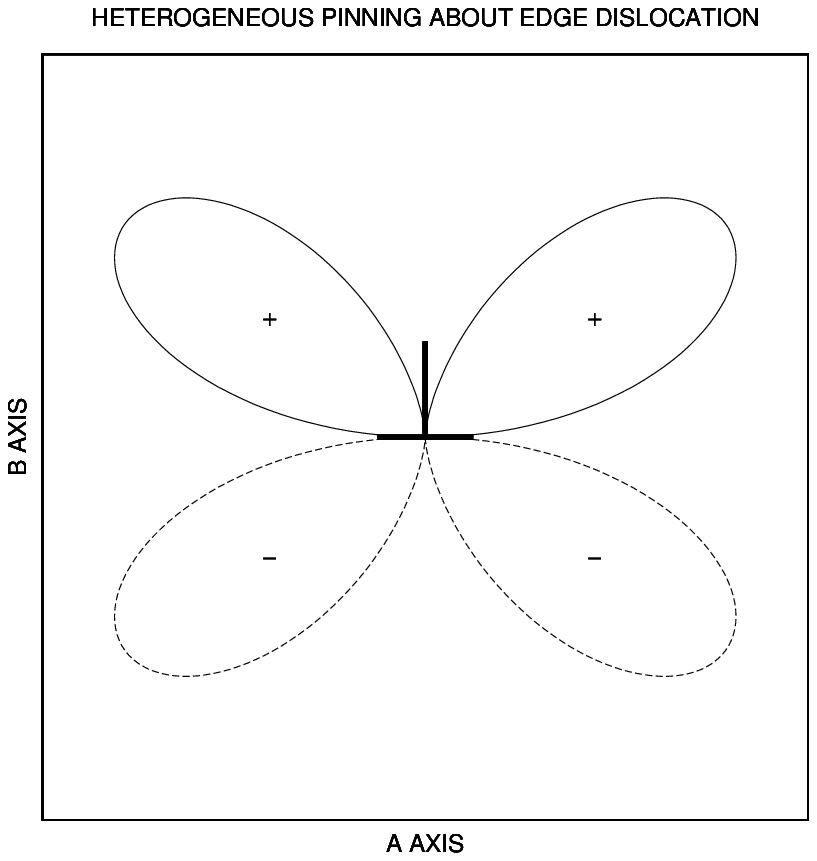}
\caption{Shown is the contour of a potential-energy landscape for a vortex
core about an edge dislocation aligned parallel to the $c$ axes
of optimally doped YBCO at two fixed energies $\pm\varepsilon_1$.}
\label{dislocation}
\end{figure}

\begin{figure}
\includegraphics[scale=1.30, angle=0]{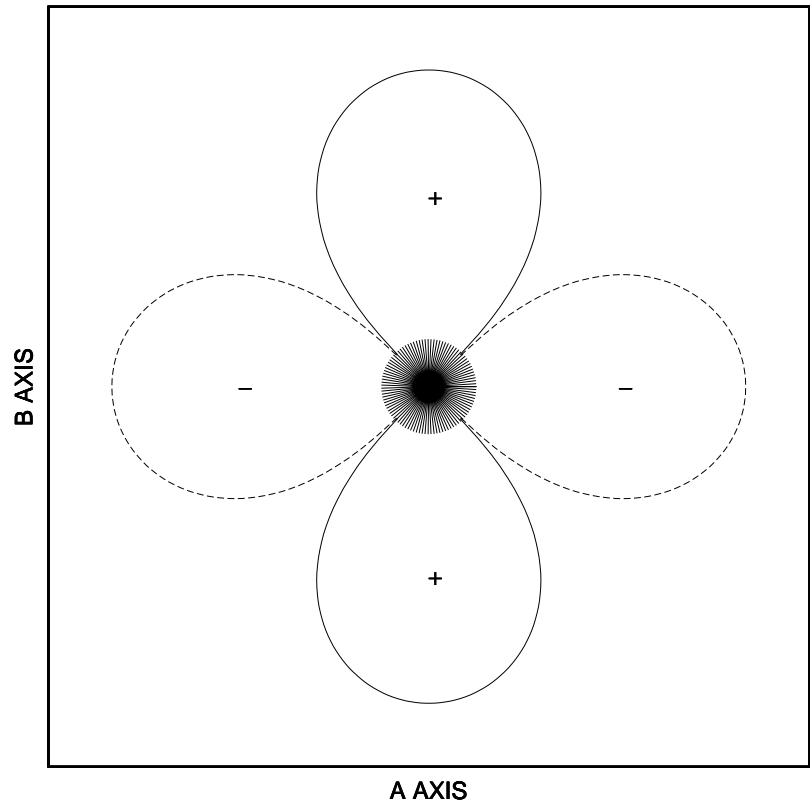}
\caption{Contours for the potential energy of a vortex core about a nano-column
inclusion in YBCO are depicted for two fixed energies $\pm\varepsilon_1$.}
\label{nano_column}
\end{figure}

In this paper, 
we shall compute the in-field critical due to natural or to artificial
columnar microstructures in YBCO thin films by employing two-dimensional (2D)
collective pinning theory\cite{kes_tsuei}\cite{jpr-maley} and 
by performing direct computer simulations of the corresponding 2D vortex 
dynamics\cite{jensen_88}\cite{kosh_vino_94}\cite{niels_96}\cite{middleton_96}\cite{ryu_96}\cite{chandran_04}\cite{chandran_05}.  
Pinning of vortex lines will be assumed to be due to variations in
$T_c$ about columnar defects that in turn are a 
result of the strain field about them\cite{welp}.
Heterogeneous columnar defects are thereby predicted 
that are composed of component lines 
of pins and of antipins of equal strength\cite{gure_pash}\cite{rbv}.
These form a dipolar ``clover-leaf'' pattern in the case of an
edge dislocation line defect (see fig. \ref{dislocation}),
while they form a quadropolar pattern in the case of
a nano-column inclusion (see fig. \ref{nano_column}).
Both 2D collective pinning theory\cite{kes_tsuei}\cite{jpr-maley}
and direct computer simulation of
2D vortex dynamics obtain the following central result:
that the critical current due to an arrangnement of
weak heterogeneous columnar pins is equal to that expected from
an arrangement of conventional columnar pins\cite{civale91}
obtained by
unbinding the component pin/antipin lines from the heterogeneous cluster,
and by then converting the isolated antipin lines that result into
pin lines of equal strength.  
Figure \ref{pin_landscapes} depicts the two pinning landscapes in question.
The antipin components of heterogeneous columnar defects therefore
increase the critical current by a factor
of two compared to the contribution due to the true pin components alone
in the collective pinning regime.

\section{Strain/pinning landscape about material line defects in YBCO}
\label{strain}
It is well known that the critical temperature in optimally doped YBCO couples
most strongly to pure shear strain in the $a$-$b$ plane\cite{welp}.  
In particular,
the variation in the critical temperature depends linearly on the 
symmetric strain tensor
$\epsilon_{\alpha , \beta }
 = (\partial_{\alpha}  u_{\beta}  + \partial_{\beta} u_{\alpha}) /2$
following
\begin{equation}
\delta T_c ({\bf r}) = \sum_{\alpha}\sum_{\beta}
(\partial{T_{c}} / \partial\epsilon_{\alpha , \beta })
\epsilon_{\alpha , \beta } ({\bf r})
\label{Tc}
\end{equation}
with strain derivatives 
$\partial{T_{c}} / \partial\epsilon_{aa} = 230\,{\rm K}$ and
$\partial{T_{c}} / \partial\epsilon_{bb} = - 220\,{\rm K}$.
Below, we shall show how the shear strain about a line
of edge dislocations or about
a nano-column inclusion in a YBCO thin film 
combined with this experimental fact
results in a heterogeneous columnar pinning line 
for a vortex line in the mixed state.

Consider a line of edge dislocations that threads a YBCO
film in parallel to the $c$ axis.  Gurevich and Pashitskii calculated the
variation in the critical temperature expected from a linear dependence on 
strain (\ref{Tc}) and obtained the result\cite{gure_pash}
\begin{equation}
\delta T_c ({\bf r}) =
\Biggl(T_{\parallel} {1-2\sigma\over{1-\sigma}}
\pm T_{\perp}{{\rm cos}^2 \phi\over{1-\sigma}}\Biggr)
 {b\, {\rm sin} \phi\over{2 \pi r}} ,
\label{disc}
\end{equation}
with constants\cite{welp}
\begin{eqnarray}
T_{\parallel} &=& -{1\over 2}\Bigl({\partial T_c\over{\partial\epsilon_{aa}}}
 + {\partial T_c\over{\partial\epsilon_{bb}}}\Bigr) 
= - 5\, {\rm K},\\
T_{\perp} &=& {\partial T_c\over{\partial\epsilon_{bb}}}
 - {\partial T_c\over{\partial\epsilon_{aa}}}
= - 450\, {\rm K}.
\end{eqnarray}
Above, $\sigma$ denotes the Poisson ratio and $b$ denotes the magnitude
of the Burgers vector. It can be aligned along either the $a$ ($+$) 
or the $b$ ($-$) axes.
We now note that the contribution of the vortex core to the vortex line tension
is approximated by the fundamental energy scale per unit length
$\varepsilon_0 = (\Phi_0/4\pi \lambda_L)^2$,
where $\lambda_L$ denotes the London penetration depth,
and where $\Phi_0$ denotes the flux quantum in a type-II superconductor.
The potential-energy landscape experienced by a  
vortex line then has a contribution
$\delta\varepsilon_1 ({\bf r}) = 
(\partial\varepsilon_0 / \partial{T_{c}}) \cdot \delta T_c ({\bf r})$
due to the variation in the critical temperature about the dislocation
line (\ref{disc}).  It is dominated by the shear component ($T_{\perp}$)
in the case of optimally doped YBCO, which is shown in fig. \ref{dislocation}.
The potential energy landscape is essentially dipolar in such case, 
with two anti-pinning lobes (+) and two pinning lobes (-).

An analogous heterogeneous pinning potential 
for a vortex line about a nano-column inclusion has been obtained more
recently by one of the authors in collaboration with 
Barnes and Varanasi\cite{rbv}.  The lattice mismatch between the inclusion
and the YBCO matrix results in pure shear strain 
about the inclusion.  These authors then ultimately find a variation of the
critical temperature in the YBCO matrix about the nano-column given by
\begin{equation} 
\delta T_c ({\bf r}) = T_{\perp} {\Delta r\over{r_{\rm out}}}
{c_{\parallel}^{({\rm in})}\over{c_{\parallel}^{({\rm in})} + {c_{\perp}^{({\rm out})}}}} 
\Bigl({r_{\rm out}\over{r}}\Bigr)^2 {\rm cos}\, 2\phi \ .
\label{nano_rod}
\end{equation}
Here $\Delta r = r_{\rm in} - r_{\rm out}$ is the mismatch between the radius
of the inclusion, $r_{\rm in}$, and the radius of the corresponding void in the
YBCO matrix, $r_{\rm out}$, while $c_{\parallel}^{({\rm in}, {\rm out})}$
and $c_{\perp}^{({\rm in}, {\rm out})}$
are the 2D compression and shear moduli, respectively.
The potential-energy landscape experienced by a
vortex line again then has a contribution
$\delta\varepsilon_1 ({\bf r}) =
(\partial\varepsilon_0 / \partial{T_{c}}) \cdot \delta T_c ({\bf r})$
due to such a variation in the critical temperature about the 
nano-column inclusion,
but with d-wave symmetry.
Figure \ref{nano_column} depicts the constant-energy contours.
Notice that the potential energy
is characterized by diametrically opposite lobes of pins and of 
antipin oriented with respect to each other by 90 degrees.

Last, the force ${\bf f}_1 ({\bf r}) = - \nabla\,\varepsilon_1 ({\bf r})$ 
that a vortex core experiences due to changes in 
$T_c$ about both a line of edge dislocations (\ref{disc}) 
and about a nanocolumn inclusion (\ref{nano_rod}) is long-range.
Suppose now that a random field of parallel lines of heterogeneous pins thread a
YBCO film along $c$ axis at transverse locations $\{ {\bf R}_i \}$.  
Statistical cancellations in the net force
experienced by a vortex core,
${\bf f} ({\bf r}) = \sum_i {\bf f}_1 ({\bf r} - {\bf R}_i)$,
cut off the long-range force. 
The effective finite range $r_{\rm p}$ of each heterogeneous pin is determined by
matching the fluctuation in the net force over the system with the long-range
force due to a single material line defect:
$\overline{f^2} = |{\bf f}_1 (r_{\rm p})|^2$.
It is given by the expression
$r_{\rm p} = (2/\pi n_{\phi})^{1/6} r_{\rm out}^{2/3}$
in the case of strain-induced pinning of vortex lines 
about nano-column inclusions\cite{rbv}, where $n_{\phi}$ denotes their density.

\section{Collective pinning of Abrikosov vortex lattice by heterogeneous columnar pins}
\label{theory}
Above, we have shown how the strain field about edge dislocations and nano-column
inclusions that thread a YBCO film along the $c$ axis result in heterogeneous
pinning lines for vortex lines that are present in the mixed phase.  
As shown by figs. (\ref{dislocation}) and (\ref{nano_column}), 
the heterogeneous columnar pins
are clusters with equal numbers of pins and antipin, each of equal strength.
[See also  Eqs. (\ref{disc}) and (\ref{nano_rod}).]
Consider now a YBCO film threaded by a field of such heterogeneous pinning lines
in applied  magnetic field aligned along the $c$ axis as well.
Next, suppose that a rigid vortex lattice appears as a result.
Among the fraction of vortex lines that experience forces from the heterogeneous
columnar pins, half will lie in valleys of the vortex-core potential 
while the other half will lie on the hills of the vortex-core potential.
Because the magnitude of the force experienced by the vortex lines that lie
in the valleys and on the hills of the vortex-core potential are equal,
the critical current should be the same as that due to a field of columnar pins
which results from 
({\it i}) unbinding the cluster of pins and antipins from the heterogeneous pins,
and ({\it ii}) converting each of the antipins that are now isolated into pins
of equal strength and range.
Figure \ref{pin_landscapes} depicts the two columnar 
pinning arrangments in question for the case of nano-column inclusions.
Below, we shall see that this prediction is indeed borne out by 2D collective
pinning theory.

Previous work by one of the authors and Maley\cite{jpr-maley} implies that
many heterogeneous columnar defects
collectively pin a rigid Abrikosov vortex lattice.
Theoretical arguments and numerical Monte Carlo simulations indicate,
in particular, that
a hexatic Bose glass state can exist at low temperature
in such a case\cite{jpr07}\cite{nono_04}.
Here the vortex lattice is threaded by
isolated lines of edge dislocations
parallel to the lines of 
relatively weak pins/anti-pins.
(See fig. \ref{critical_state_config}.)
The critical current $j_c$ is then limited by
plastic creep of the vortex lattice
associated with glide by such edge dislocations\cite{jpr-maley}.
Application of 2D collective pinning theory\cite{kes_tsuei}
yields a transverse Larkin scale $R_c$ that is set by the density of Larkin domains
\begin{equation}
R_c^{-2} \sim n_{\rm p} (f_{\rm p} / c_{66} b_{\triangle})^2.
\label{R_c}
\end{equation}
Here $n_{\rm p}$ denotes the density of vortex lines 
pinned by the material line defects
and $f_{\rm p}$ denotes the maximum force per unit length
at a pin/antipin that makes up the heterogeneous material line defect. 
Also, $c_{66}$ denotes the shear modulus of the vortex lattice,
while $b_{\triangle}$ is the magnitude of Burgers vectors that glide.
It can be shown that $R_c$ is of order the separation between isolated lines of
dislocations in the hexatic Bose glass\cite{M-E}.
A statistical summation of the pinning forces inside of a corresponding Larkin volume,
in which the vortex lattice is perfectly rigid,
then yields a net pinning force density
\begin{equation}
j_c  B / c  \sim n_{\rm p} f^2_{\rm p} / c_{66} b_{\triangle}.
\label{critical_state}
\end{equation}
The above is valid if many vortex lines experience pinning forces inside of a Larkin
volume: $n_{\rm p} R_c^2 \gg 1$.  
Last, study of 
equations (\ref{disc}) and (\ref{nano_rod}) shows that
the potential energy
about a heterogeneous pinning line experienced by a vortex line
has  zero angle average.
This implies that the occupation of a heterogeneous pin by
a vortex line is purely random.
The density of vortex lines that they collectively pin
is then equal to
$n_{\rm p} = (\sigma_{\rm p} n_B) n_{\phi}$, where
$n_{\phi}$ denotes the density of heterogeous columnar defects, 
where
$\sigma_{\rm p} = \pi (r_{\rm p}^2 - r_{\rm out}^2)$
is the effective crossectional area of a heterogeneous  pinning/antipinning line,
and where $n_B$ denotes the density of vortex lines.
Observe now that both $f_{\rm p}$ and the product $n_{\phi} \sigma_{\rm p}$
remain constant after unbinding the pins and anti-pins that make up heterogeneous
columnar defects, and after then converting the isolated antipins that result
into isolated true pins!  
We conclude that 2D collective pinning theory\cite{jpr-maley}\cite{kes_tsuei} implies that
the critical current due to comparable arrangements of 
conventional columnar pins and of heterogeneous columnar pins
like those shown in fig. \ref{pin_landscapes} are the same.

\section{Computer simulation of 2D vortex dynamics}
\label{lngvn}
We shall now compute the in-field critical current due to columnar pins
instead by
simulating over-damped vortex dynamics with the Langevin equation. 
Our aim is to once again
compare the  critical-state performance of type-II superconductors threaded by 
conventional columnar pins\cite{civale91} with that of
type-II superconductors threaded by
heterogeneous (d-wave) pinning lines that can result from the strain field about
nano-column inclusions in YBCO\cite{rbv}.  
The Abrikosov vortex lattice in a field
of parallel columnar pins is in a Bose glass state characterized by
an infinite macroscopic tilt modulus\cite{N-V}.
The longitudinal Larkin scale $L_c$ is then infinite\cite{jpr07}, 
which means that vortex lines may be considered as rigid rods 
at long longituginal length scales.
In a film geometry that is perpendicular the applied magnetic field,
their dynamics can therefore be described by 
the over-damped 2D Langevin equations for the transverse motion
of a given rigid vortex line at location ${\bf r}$ 
\cite{jensen_88}\cite{kosh_vino_94}\cite{niels_96}\cite{middleton_96}\cite{ryu_96}\cite{chandran_04}\cite{chandran_05}:
\begin{equation}
\eta \dot{\bf r} = \sum_{{\bf r}^{\prime}}{\bf F}_{vv} + 
{\bf F}_{vp} + {\bf F}_{\rm Lorentz} + {\bf F}_{\rm Brown}(t). 
\label{langevin}
\end{equation}
It is governed by the Bardeen-Stephen viscosity $\eta$ for a vortex line.
Each rigid vortex line experiences forces
${\bf F}_{vv} = -{\bf\nabla} U_{vv}$ and ${\bf F}_{vp} = -{\bf\nabla} U_{vp}$
due to interactions with other vortices
at locations ${\bf r}^{\prime}$
and due to the field of material line defects, respectively. 
The former is taken to be long range:
\begin{equation}
U_{vv}({\bf r}, {\bf r}^{\prime}) =
 {4\pi\varepsilon_0 d\over{N_x N_y}} \sum_{{\bf q}\neq 0}
{{\rm exp}[{\bf q}\cdot ({\bf r} - {\bf r}^{\prime})]\over{4 - 2\, {\rm cos}(q_x a) - 2\, {\rm cos}(q_y a)}}.
\label{interaction}
\end{equation}
Note that periodic boundary conditions are imposed over a rectangle of dimensions
$N_x a\times N_y a$.  The lattice constant $a$ of the grid is understood to be
of order the diameter of a vortex core.
Above, $d$ denotes the thickness of the film.
Values for both potential energies
$U_{vv}$ and $U_{vp}$ that span a $N_x\times N_y$
grid are stored in look-up tables. 
The corresponding forces ${\bf F}_{vv}$ and ${\bf F}_{vp}$ 
are computed from these potentials by taking finite differences over the grid.
They are also stored in look-up tables.
Last, a force at an arbitrary location is determined by 
linear interpolation from those determined off the grid.

\begin{figure}
\includegraphics[scale=1.00, angle=0]{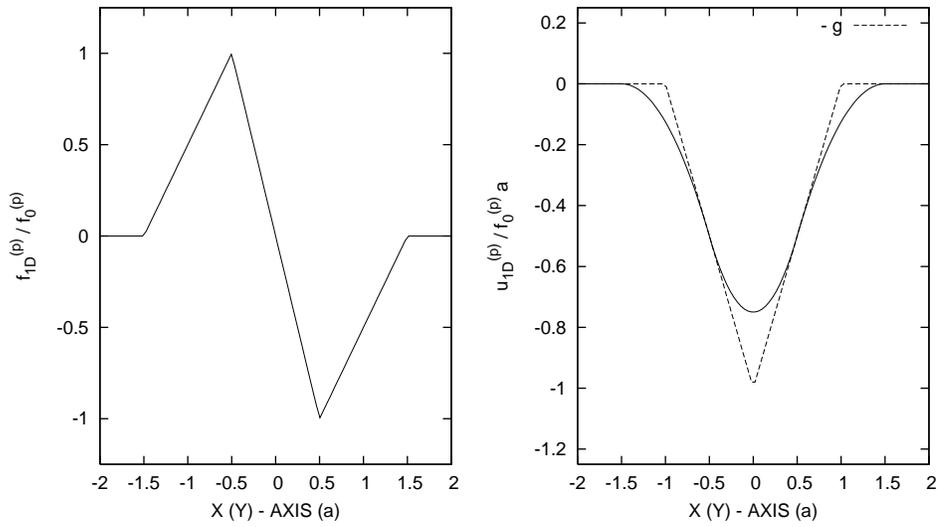}
\caption{The solid lines above depict the pinning force and the 
corresponding potential energy along principal axes ($x=0$ or $y=0$)
due to a featureless point defect that lies on the square lattice
grid at point $(0,0)$.  The force at a general point $(x,y)$ is 
determined by linear interpolation over the grid.}
\label{conservative}
\end{figure}

In the special case of a featureless (S-wave)
line pin that lies at a grid point $(ma,na)$,
the above linear interpolation scheme yields a pinning force of the form
\begin{eqnarray*}
f_x^{({\rm p})} (x,y) &=& f_{1D}^{({\rm p})}(x-ma) \cdot g(y-na),\\
f_y^{({\rm p})} (x,y) &=& g(x-ma) \cdot f_{1D}^{({\rm p})}(y-na),
\end{eqnarray*}
where $f_{1D}^{({\rm p})}$ is the pinning force along a principal axis of the grid that contains
the pinning center.  Both it and its corresponding 1D potential energy 
$u_{1D}^{({\rm p})}$ are shown in fig. \ref{conservative}.  
The scale $f_0^{({\rm p})}$ that appears there
denotes the magnitude of the maximum pinning force,
which is related to the depth $u_{0}^{({\rm p})}$ of the line pin potential by
$f_{0}^{({\rm p})} = u_{0}^{({\rm p})}/a$.
The transverse interpolation of forces is set by the function $g$,
which is also shown in fig. \ref{conservative}.  
It is given explicitly by $g(y) = 1 - |y/a|$ if $|y| < a$, 
and by $g(y) = 0$ otherwise.
Notice now that
$\partial_x f_y^{({\rm p})} \neq \partial_y f_x^{({\rm p})}$,
which means that the above force field for a line pin is not conservative.
On the other hand, 
leaving the 1D function $g$ free while imposing a conservative force
implies that it must be given by 
$g(y) = - u_{1D}^{({\rm p})}(y) / f_{0}^{({\rm p})} a$ instead.  
As shown by fig. \ref{conservative},
both functions have qualitatively similar shapes. 
They both also have the same area,
$\int_{-\infty}^{+\infty} dy\, g(y) = a$.
We believe, therefore, that the model for a linear pin displayed
by the equations above is qualitatively correct.

\begin{figure}
\includegraphics[scale=1.30, angle=0]{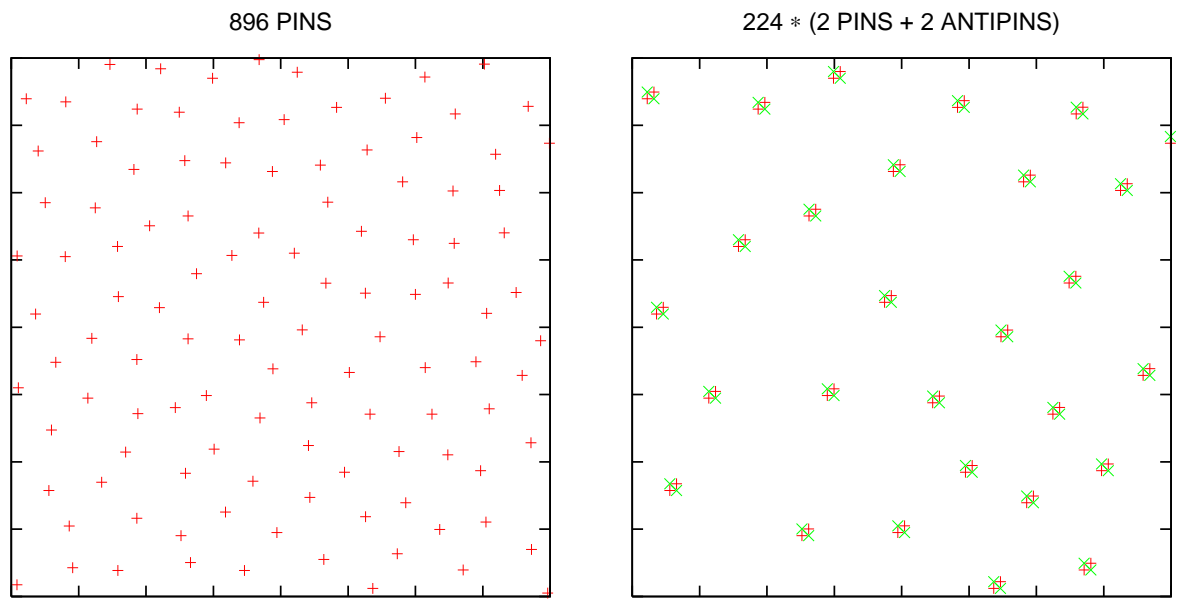}
\caption{Shown above are windows of equal area into
an arrangement of conventional columnar pins ($+$)
and into an arrangment of quadrapolar columnar pins ($+ \times , \times +$) 
that can result from shear strain about nano-column inclusions.}
\label{pin_landscapes}
\end{figure}

The rigid vortex lines also experience a uniform Lorentz force 
${\bf F}_{\rm Lorentz}$
that push them across the width of the wire, 
in addition to a random Brownian force 
${\bf F}_{\rm Brown}(t)$ that brings the vortex lattice into thermal equilibrium.
We numerically evolved the above Langevin equation (\ref{langevin}) 
in order to describe the over-damped dynamics of $896$ rigid vortex lines.
The second-order Runge-Kutta method was employed.
Values for the interaction forces and for the pinning forces
to be stored in look-up tables  
were determined by taking finite differences 
over a $224 \times 224$ square grid that spans the system,
and which has a lattice constant $a$ that is  understood to 
be of order the superconducting coherence length.
The computer simulations were conducted in two stages.  
We first annealed the vortex lattice down to low temperature 
in the presence of a 
landscape of weak columnar pins.
Figure \ref{pin_landscapes} depicts the two pinning (anti-pinning) landscapes
that were studied.
All (component) line pins had equal strengths,
each with a maximum pinning/antipinning energy 
$u_{0}^{({\rm p})} = f_{0}^{({\rm p})} a$ 
equal to 1\% of the dominant interaction scale $2\varepsilon_0 d$.
The initial temperature was taken to be very 
many times the Halperin-Nelson-Young melting
temperature of the 2D  vortex lattice, $T_m^{(2D)}$, 
and it was successively halved 
until a very small fraction of $T_m^{(2D)}$ was reached.
Two million time steps were taken at each temperature, 
with a time interval $\Delta t$
equal to a few percent of the natural time
scale $\eta a_v^2 / 4\pi \varepsilon_0 d$
that is set by the dominant interaction energy among vortex lines,
Eq. (\ref{interaction}). 
Here $a_v = (\Phi_0 / B)^{1/2}$ is the average separation between
vortex lines.  
The annealed vortex lattice was found to be in a hexatic vortex
glass state characterized by long-range orientational order 
and unbound dislocations\cite{jpr07}\cite{nono_04}\cite{chudnovsky_91}.
Only a fraction of the vortices in the lattice
experience pinning forces.
In the second stage, we drove the annealed vortex lattice 
at zero temperature with a uniform Lorentz
force in the presence of the same landscape of columnar pins.  
The driving force increased in 300 steps from zero to about twice the critical force (current).
Each step of constant Lorentz force lasted $200,000$ time iterations, $\Delta t$,
and the Lorentz force was increased at each step 
by a part in $100,000$ of the interaction scale $2\varepsilon_0 d / a_v$.

\begin{figure}
\includegraphics[scale=1.30, angle=0]{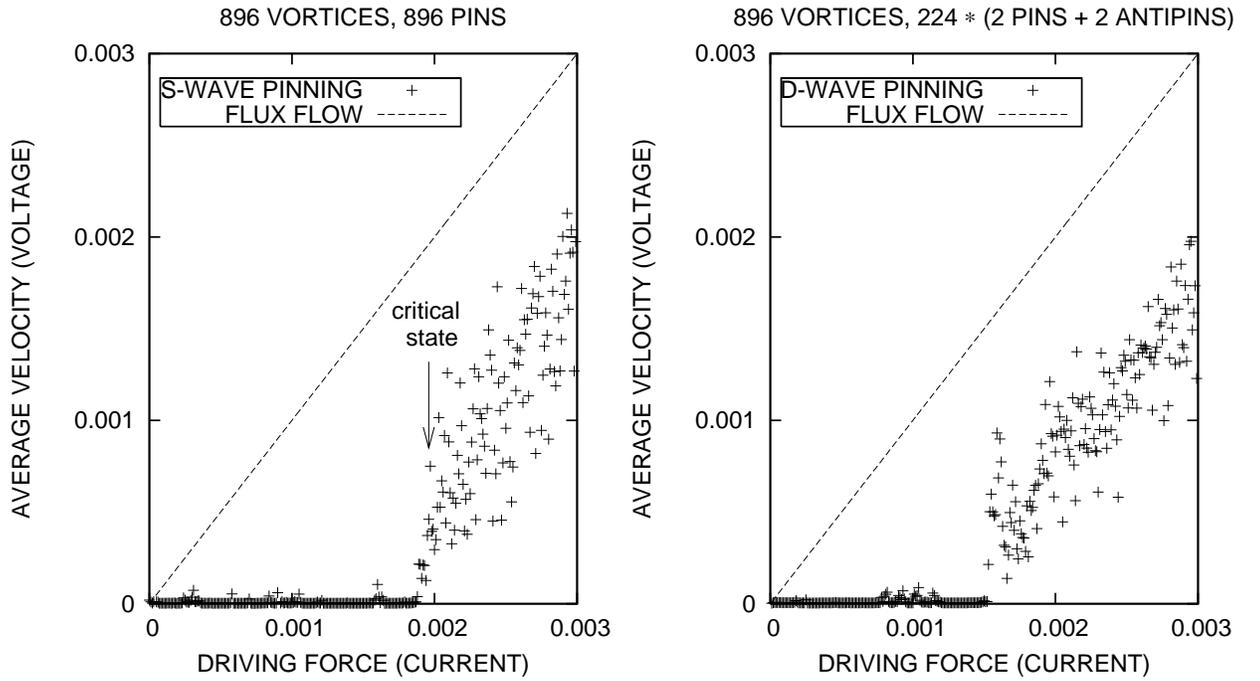}
\caption{The left panel above shows the current-voltage curve obtained from
driving 896 rigid vortex lines in the presence of an equal number of
conventional (S-wave) columnar pins via the Langevin equation (\ref{langevin}).
All forces and velocities are given in units of 
$2\varepsilon_0 d / a_v$ and of $2\varepsilon_0 d / \eta a_v$,
respectively.
The right panel above depicts the current voltage curve obtained from
driving 896 rigid vortex lines in the presence of 
224 quadrapolar  columnar pins.
Each component pin/antipin line has a strength and range equal to
that of the conventional S-wave columnar pins.
(See fig. \ref{pin_landscapes}.)}
\label{Jc}
\end{figure}

Figure \ref{Jc} shows the  current-voltage properties
that result from driving the 
vortex lattice at zero-temperature through the pinning landscapes  
that are shown in fig. \ref{pin_landscapes}.
The landscape in the left panel is the field of
conventional (S-wave) columnar pins that is shown in the
left panel of fig. \ref{pin_landscapes}.
All of the pins in this case are featureless and of equal strength,
with a number that matches the number of vortex lines (896).
As depicted by fig. \ref{conservative},
each pin lies at a grid point 
and inherits the minimum range of the grid.
Notice  the  critical current that separates
a pinned vortex lattice at low currents from a vortex lattice
that experiences flux creep and flux flow at high currents.  
It is interesting to note that although the vortex lattice 
was threaded by isolated dislocation defects throughout the simulation,
these became well seperated precisely at the critical state.
(See figs. \ref{critical_state_config} and \ref{Jc}.)
The dislocation defects then gradually began to pair up and annihilate 
with increasing driving force in the flux-creep/flux-flow regime.
The right panel of fig. \ref{Jc} displays the current-voltage property for
a driven vortex lattice that experiences
the arrangement of quadrapolar columnar pins that is shown in the right
panel of fig. \ref{pin_landscapes}.
The latter aims to model the strain-induced pinning potential 
shown in fig. \ref{nano_column}
that is predicted to exists about
nano-column inclusions in YBCO films\cite{rbv}.
This pinning landscape contains
half the number ($2\times 224$) of true line pins,
 by comparison with the previous case,
with an equal number ($2\times 224$) of antipin lines.
Each component line of pins/antipins has the same range and strength 
as in the previous case,
on the other hand.
Figure \ref{Jc} indicates that the  critical current
in the present case of heterogeneous columnar pins 
is comparable to that obtained in the previous case of featureless columnar pins,
which had {\it twice} the number of true pins!  
This result is  nevertheless anticipated by 2D collective pinning theory,
Eq. (\ref{critical_state}).
It is due to the fact that the pin and the antipin components of the 
heterogeneous pinning line contribute equally to 
the critical current of a rigid vortex lattice.
Last, two well-spaced dislocation defects
with equal and opposite Burgers vectors
threaded the vortex lattice throughout the second simulation.

\section{Conclusions}
\label{conclude}
\begin{figure}
\includegraphics[scale=0.80, angle=0]{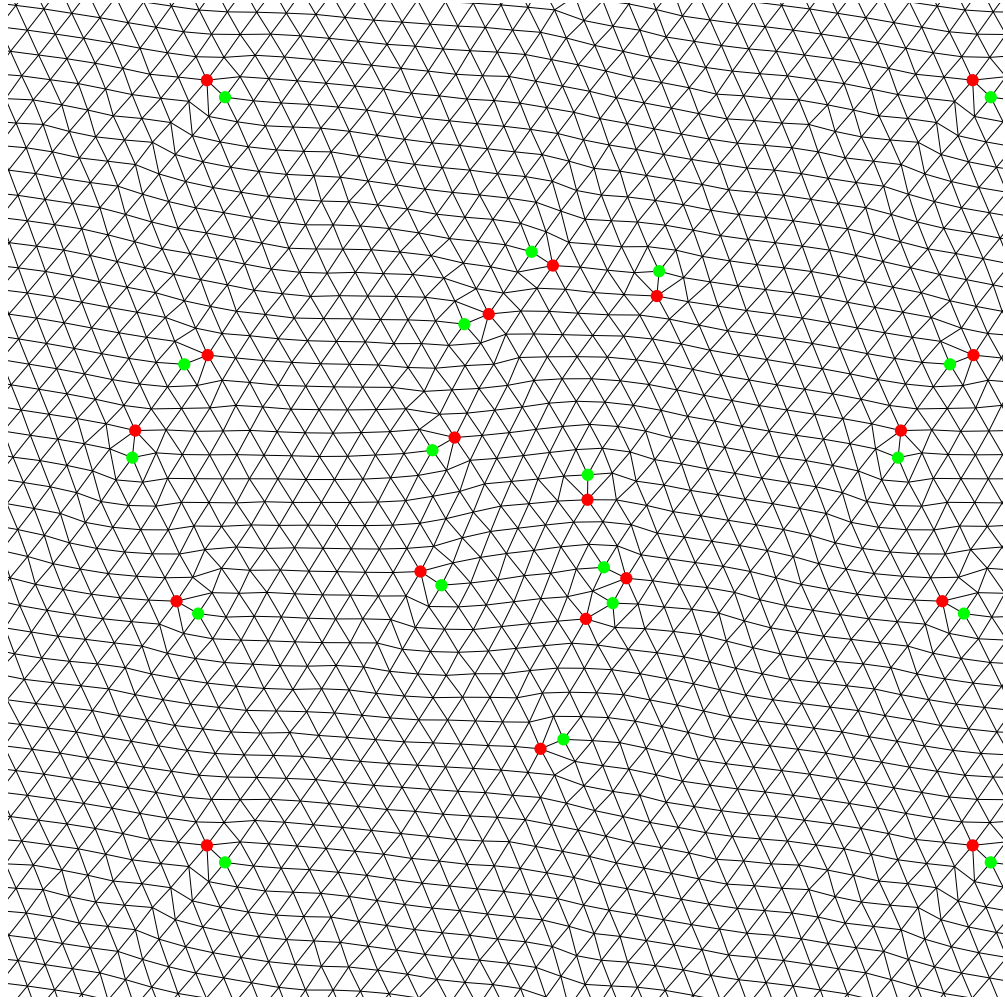}
\caption{Shown is the critical-state
configuration of 896 rigid vortex lines in the presence of 
an equal number of conventional columnar pins.  
(See the left panel in fig. \ref{Jc}.)
Pairs of red and green dots mark dislocation defects in the triangular
vortex lattice.}
\label{critical_state_config}
\end{figure}

In summary, strain about columnar microstructure in YBCO films
generates heterogeneous pin/antipin lines that can multiply the 
critical current expected from the 
true pin line components alone by a factor of two.
This result is  obtained for weak correlated pinning 
inside of the collective pinning regime, however.  
It is unlikely to persist in the limit of
strong correlated heterogeous pins,
where all vortex lines that experience pinning forces then
lie along a true pin component of the composite line defect.
Antipin line components of the heterogenous
pin have no effect on the critical current in that limit.
A key assumption of the collective pinning analysis
is that the critical state be found in a hexatic Bose glass state characterized 
by isolated dislocation defects that thread the vortex lattice\cite{jpr-maley}. 
Figure \ref{critical_state_config} demonstrates that
this assumption is indeed confirmed by
the molecular dynamics simulations that we conducted across the critical state 
for 896 rigid vortex lines.
Previous 2D simulations find evidence for the break-up of the vortex lattice
into grains at much higher numbers of rigid vortex lines when the latter
experience short-range interactions\cite{chandran_04}\cite{chandran_05}.
Thermodynamic Monte Carlo simulations on
over 2000 vortices that experience the 
same long-range logarithmic interaction  studied here, Eq. (\ref{interaction}),
find evidence for an equilibrium hexatic vortex glass phase\cite{jpr07},
on the other hand.
It therefore remains to be seen if the
hexatic glass phase obtained in the present
molecular-dynamics simulations persists in the thermodynamic limit.

JPR thanks Paul Barnes and Hsin-Ju Wu for discussions.  
This work was supported in part by the US Air Force
Office of Scientific Research under grant no. FA9550-06-1-0479.



\end{document}